\documentstyle[multicol,aps,epsfig]{revtex} 

\tighten       

\begin{document}
\draft 

\title{Microscopic derivation of transport coefficients and 
boundary conditions in discrete drift-diffusion models of
weakly coupled superlattices}
\author{ L. L. Bonilla }
\address{Departamento de Matem\'{a}ticas, Escuela Polit\'ecnica
Superior,  Universidad Carlos III de Madrid,\\
Avenida de la Universidad 30, 28911 Legan{\'e}s, Spain\\
Also: Unidad Asociada al Instituto de Ciencia de Materiales
(CSIC)}
\author{G. Platero and D. S\'anchez}
\address{Instituto de Ciencia de Materiales (CSIC),
Cantoblanco, 28049 Madrid, Spain}
\date{\today}
\maketitle
\begin{abstract}
A discrete drift-diffusion model is derived from a microscopic sequential
tunneling model of charge transport in weakly coupled superlattices
provided temperatures are low or high enough. Realistic transport
coefficients and novel contact current--field characteristic
curves are calculated from microscopic expressions, knowing the
design parameters of the superlattice. Boundary conditions clarify
when possible self-sustained oscillations of the current are due
to monopole or dipole recycling. 
\end{abstract}

\begin{multicols}{2}
\narrowtext

\setcounter{equation}{0}
\section{Introduction}
\label{sec:intro}
At present, the theory of charge transport and pattern formation in
superlattices (SL) is in a fragmentary state. On the one hand, it is
possible to establish a quantum kinetic theory from first principles
by using Green function formalisms \cite{kad62,hau96}. However,
the resulting equations are hard to solve, even numerically, unless
a number of simplifications and assumptions are made \cite{hau96,wac98}.
These include: (i) a constant electric field, (ii) simplified
scattering models, and (iii) a stationary current through the SL.
These assumptions directly exclude the description of electric field
domains and their dynamics although important results are still
obtained \cite{wac98,wac99}. The stationary current density probes the
difference between strongly and weakly coupled SL. It also indicates
when simpler theories yield good agreement with quantum kinetics.
The main simpler theories are (see Figure 1 of Ref.\ \onlinecite{wac98}): \\
(i) Semiclassical calculations of miniband transport using the
Boltzmann transport equation \cite{pal92} or simplifications thereof, such
as hydrodynamic \cite{but77} or drift-diffusion \cite{sib89} models. These
calculations hold for strongly coupled SL at low fields. In the miniband
transport regime, electrons traverse the whole SL miniband thereby
performing Bloch oscillations and giving rise to negative differential
conductivity (NDC) for large enough electric fields \cite{esa70}.
The latter may cause self-sustained oscillations of the current due to
recycling of charge dipole domains as in the Gunn effect of bulk n-GaAs
\cite{gun63,but77}. \\
(ii) Wannier-Stark (WS) hopping transport in which electrons move parallel to
the electric field through scattering processes including hopping transitions
between WS levels \cite{tsu75}. Calculations in this regime hold for
intermediate fields, larger than those corresponding to collisional
broadening of WS levels, but lower than those corresponding to resonant
tunneling. \\
(iii) Sequential tunneling calculations valid for weakly coupled SL
(coherence length smaller than one SL period) at basically any value
of the electric field \cite{kaz72,agu97,wac97a}. A great advantage
of this formulation as compared with (i), (ii) or Green function
calculations is that boundary conditions can be derived naturally
and consistently from microscopic models \cite{agu97}. 

On the other hand, the description of electric field domains and
self-sustained oscillations in SL has been made by means of discrete
drift models. These models use simplified forms of the tunneling current
through SL barriers and discrete forms of the charge continuity and
Poisson equations \cite{pren94,bon94,bon95}. Although discrete
drift models yield good descriptions of nonlinear phenomena in SL,
bridging the gap between them and more microscopic descriptions
\cite{agu97,wac97a} is greatly desirable for further advancing both
theory and experiments.

A step in this direction is attempted in the present paper. Our
starting point is a microscopic description of a weakly coupled SL
by means of discrete Poisson and charge continuity equations. In
the latter the tunneling current through a barrier is a function
of the electrochemical potentials of adjacent wells and the
potential drops in them and in the barrier. This function is
derived by means of the Transfer Hamiltonian method provided the
intersubband scattering and the tunneling time are much smaller
than the typical dielectric relaxation time \cite{agu97}. From this
microscopic model and for sufficiently low or high temperatures, we
derive discrete drift-diffusion (DDD) equations for the field and
charge at each SL period and appropriate boundary conditions. The
drift velocity and diffusion coefficients in the DDD equations are
nonlinear functions of the electric field which can be calculated
from first principles for any weakly coupled SL. These equations
are of great interest for the study of nonlinear dynamics in SL.
They are simpler to study than microscopic model equations for
which only numerical simulation results are available
\cite{san99}. 

In the present work, natural boundary conditions for DDD equations
are derived from microscopic calculations for the first time.
Previous authors had to propose boundary conditions with adjustable
parameters which gave qualitative agreement with experimental
results \cite{wac97a,pren94,bon94,bon95,kas97}. The present 
boundary conditions relate current density and field at contacts
and can be calculated for a given configuration of emitter and
collector contact regions. As it is well-known, boundary conditions
select the stable charge and field profiles in the SL, and
therefore are crucial to understand which spatio-temporal
structures will be observed in the SL for given values of the
control parameters \cite{wac97a,pren94,bon94,bon95,kas97}.  

The rest of the paper is as follows. In Section \ref{sec:2}, we
review the microscopic sequential resonant tunneling model. We 
obtain the minimal set of independent equations and boundary
conditions describing this model. Our derivation of the DDD model
is presented in Section \ref{sec:3}. Numerical evaluation of
velocity, diffusion and contact coefficients for several SLs is
presented in Section \ref{sec:4}. Section \ref{sec:5} contains our
conclusions. The Appendix contains an evaluation of the transport
coefficients for negative values of the electric field.

\section{Microscopic sequential tunneling model}
\label{sec:2}
The main charge transport mechanism in a weakly coupled SL is
sequential resonant tunneling. We shall assume that the macroscopic
time scale of the self-sustained oscillations is larger than the
tunneling time (defined as the time an electron needs to advance
from one well to the next one). In turn, this latter time is
supposed to be much larger than the intersubband scattering time.
This means that we can assume the process of tunneling across
a barrier to be stationary, with well-defined Fermi-Dirac
distributions at each well, which depend on the instantaneous
values of the electron density and potential drops. These densities
and potentials vary only on the longer macroscopic time scale.

\subsection{Tunneling current }
\begin{figure}
\centerline{\epsfig{file=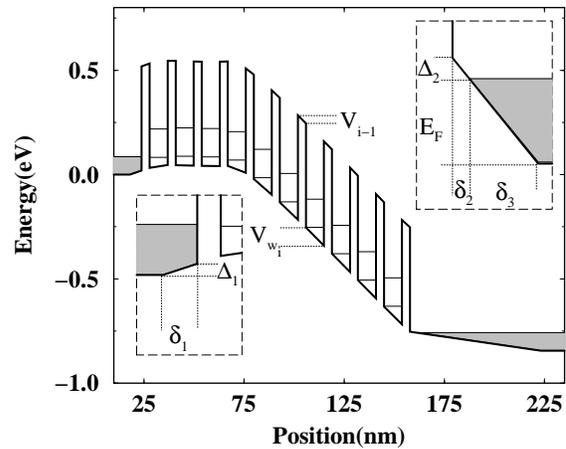,angle=270,width=0.45\textwidth}}
\vspace{0.5 cm} 
\caption{ Sketch of the electrostatic potential profile in a SL.}
\label{fig1}
\end{figure}

The tunneling current density across each barrier in the SL may
be approximately calculated by means of the Transfer Hamiltonian
method. We shall only quote the results here \cite{agu97}. Let
$e J_{e,1}$ and $e J_{N,c}$ be the currents
in the emitter and collector contacts respectively, and let
$e J_{i,i+1}$ be the current through the $i$th barrier which
separates wells $i$ and  $i+1$. We have
\begin{eqnarray}
J_{e,1}&\equiv & J_{0,1} = \frac{k_{B}T}{2\pi^{2}\hbar}
\sum_{j=1}^{n} \int A_{Cj}^{1}(\epsilon)\, B_{0,1}(\epsilon)\,
T_{0}(\epsilon) \,\,\,\, \nonumber\\
&\times&
\ln \left[\frac{1+e^{\frac{\epsilon_{F}-\epsilon}{k_{B}T}}}
{1+e^{\frac{\epsilon_{w_{1}}-\epsilon}{k_{B}T}}}\right]\,
d\epsilon,\label{Je}\\
J_{i,i+1}& = &\frac{\hbar k_{B}T}{2\pi^{2}m^{*}}\sum_{j=1}^{n}
\int A_{C1}^{i}(\epsilon)\, A_{Cj}^{i+1}(\epsilon)\,
B_{i-1,i}(\epsilon)  \,\,\nonumber\\
&\times& B_{i,i+1}(\epsilon)
T_{i}(\epsilon)\, \ln
\left[\frac{1+e^{\frac{\epsilon_{w_{i}}-\epsilon}{k_{B}T}}}
{1+e^{\frac{\epsilon_{w_{i+1}}-\epsilon}{k_{B}T}}} \right]
d\epsilon,  \label{THM}\\
J_{N,c}&\equiv & J_{N,N+1} =
\frac{k_{B}T}{2\pi^{2}\hbar} \int A_{C1}^{N}(\epsilon)\,
B_{N-1,N}(\epsilon)   \,\,\nonumber\\  &\times&
T_{N}(\epsilon)\,  \ln
\left[\frac{1+e^{\frac{\epsilon_{w_{N}}-\epsilon}
{k_{B}T}}}{1+e^{\frac{\epsilon_{F}-eV-\epsilon}{k_{B}T}}} \right]\,
d\epsilon. \label{Jc}
\end{eqnarray}
In these expressions:
\begin{itemize}
\item $i=1,\ldots,N-1$, $n$ is the number of
subbands in each well $i$ with energies $\epsilon_{Cj}^{i}$
(measured with respect to the common origin of potential drops:
$\epsilon=0$ at the bottom of the emitter conduction band).
$\epsilon_{F}=\hbar^2 (3 \pi^2 N_D)^{{2\over 3}}/(2m_w^{*})$
are the Fermi energies of the emitter and collector regions
calculated as functions of their doping density $N_D$.
$m_{w}^{*}$ and $m^{*}$ are the effective masses of the electrons
at the wells and barriers, respectively. 
\item $B_{i-1,i}$ are given by
\begin{eqnarray}
B_{i-1,i} &=& k_{i}\,\left[w + \left({1\over\alpha_{i-1}} +
{1\over\alpha_{i}}\right)\right.\nonumber\\
&\times &\left.\left({m^{*}\over m_{w}^{*}} \sin^{2} 
{k_{i}w\over 2} + \cos^{2} {k_{i}w\over 2}\right)\right]^{-1}
\,,\label{B}\\
\hbar k_{i} &=&\sqrt{2 m_{w}^{*}(\epsilon + e W_i)}\,,\label{ki}\\
\hbar\alpha_{i} &=& \sqrt{2 m^{*} e\left[V_{b} -  W_{i} -
{V_{w_{i}}\over 2} - {\epsilon\over e} \right]}\,,\label{ai}\\
W_i &\equiv &\sum_{j=0}^{i-1}(V_{j} + V_{w_{j}}) + {V_{w_{i}}\over
2}\,,
\label{Wi}
\end{eqnarray}
where $k_{i}$ and $\alpha_{i}$ are the wave vectors in the wells
and the barriers, respectively. $k_{i}$ depends on the electric
potential at the center of the $i$th well, $W_i$, whereas
$\alpha_{i}$ depends on the potential at the beginning of the $i$th
barrier, $W_i + V_{w_{i}}/2$. See figure \ref{fig1}. $V_i$ and
$V_{w_{i}}$, $i=1,\ldots,N$, are the potential drops at the $i$th
barrier and well, respectively. We assume that the potential drops
at barrier and wells are non-negative and that the electrons are
singularly concentrated on a plane located at the end of each well
(which is consistent with the choice of $\alpha_{i}$; the choice
of $k_i$ is dictated by the Transfer Hamiltonian method). The
potential drops $V_0$ and $V_N$ correspond to the barriers
separating the SL from the emitter and collector contacts,
respectively. $\Delta_1 \equiv e V_{w_{0}} = 2 e W_0$ is the
energy  drop at the emitter region, and $e V_b$ is the barrier
height in the absence of potential drops.  
\item $T_i$ is the dimensionless transmision probability through
the $i$th barrier separating wells $i$ and $i+1$: 
\begin{eqnarray}
T_{i}(\epsilon) = {16 k_{i}k_{i+1} \alpha_{i}^{2} e^{-2
\alpha_{i}d}\over \left[k_{i}^{2} + \left({m_{w}^{*}
\alpha_{i}\over m^{*}}\right)^{2} \right]
\left[ k_{i+1}^{2} + \left({m_{w}^{*}\alpha_{i} \over
m^{*}}\right)^{2}\right]}\,, 
\label{Ti}
\end{eqnarray}
provided $\alpha_i d\gg 1$.
\item $w$ and $d$ are the widths of wells and barriers
respectively.
\item Scattering is included in our model by means of Lorentzian
functions: 
\begin{eqnarray}
A_{Cj}^{i}(\epsilon) = {\gamma\over (\epsilon -
\epsilon_{Cj}^{i})^{2} +\gamma^{2}} \label{lorenz}
\end{eqnarray}
(for the $i$th well). The Lorentzian half-width is $\gamma
=\hbar/\tau_{sc}$, where $\tau_{sc}$ is the lifetime associated to
any scattering process dominant in the sample (interface
roughness, impurity scattering, phonon scattering\ldots)
\cite{Weil,Ramon,Ramona,Glo}. For the samples considered here,
$\gamma$ ranges from 1 to 10 meV\cite{kas97}. Of course this
phenomenological treatment of scattering could be improved by
calculating microscopically the self-energy associated to one of
the scattering proccesses mentioned above. However this
restriction to one scattering mechanism would result in a loss of
generality and simplicity of the model.
\item The integration variable $\epsilon$ takes on values from the
bottom of the $i$th well to infinity.
\end{itemize}

Of course this model can be improved by calculating microscopically
the self-energies, which could include other scattering mechanisms
(e.g.\ interface roughness, impurity effects \cite{wac99,wac97a})
or even exchange-correlation  effects (which affect the electron
charge distribution in a self-consistent way). We have assumed
that the electrons at each well are in local equilibrium with
Fermi energies $\epsilon_{w_{i}}$, which define the electron
number densities $n_i$:
\begin{eqnarray}
n_i(\epsilon_{w_{i}}) = \frac{m_{w}^{*} k_{B}T}{\pi^{2}
\hbar^{2}} \int A_{C1}^{i}(\epsilon) \ln\left[1+e^{\frac{
\epsilon_{w_{i}}-\epsilon}{k_{B}T}}\right]\, d\epsilon .
\label{n(e)}
\end{eqnarray}
Notice that the complicated dependence of the wave vectors $k_i$
and $\alpha_i$ with the potential, $W_{i}$, may be transferred to
the Fermi energies by changing variables in the integrals of the
system (\ref{THM}) so that the lower limit of integration (the 
bottom of the $i$th well) is zero: $\epsilon' = \epsilon + e\,
W_i$. Then the resulting expressions have the same forms as
Equations (\ref{THM}) and (\ref{n(e)}) if $\epsilon_{C1}^{i}$,
$\epsilon_{Cj}^{i+1}$, and
$\epsilon_{w_{i}}$ in them are  replaced by
\begin{eqnarray}
\epsilon_{C1} = \epsilon_{C1}^{i} + e\, W_i ,\label{levels}\\
\mu_i \equiv \epsilon_{w_{i}} + e\, W_{i},\label{chempot}
\end{eqnarray}
respectively. $W_i$ is given by (\ref{Wi}). The integrations now
go from $\epsilon'=0$ to infinity. Notice that $\epsilon_{Cj}$ is
independent of the well index $i$ provided we assume that the
energy level drops half the potential drop for the whole well $e
 V_{w_{i}}$ with respect to its position in the absence of bias. 
Eq.\ (\ref{n(e)}) becomes
\begin{eqnarray}
n_i(\mu_{i}) = \frac{m_{w}^{*} k_{B}T}{\pi^{2}\hbar^{2}}
\int_0^{\infty} A_{C1}(\epsilon)\, \ln
\left[1+e^{\frac{\mu_{i}-\epsilon}{k_{B}T}}\right]\, d\epsilon .
\label{n(mu)}
\end{eqnarray}
Here $A_{C1}(\epsilon)$ is obtained by substituting $\epsilon_{C1}$
(the energy of the first subband measured from the bottom of a
given well, therefore independent of electrostatics) instead of
$\epsilon_{C1}^i$ in (\ref{lorenz}). Notice that (\ref{n(mu)})
defines a one-to-one relation between $n_i$ and $\mu_i$ which is
independent of the index $i$ or the potential drops. The inverse
function
$$ \mu_i = \mu(n_i,T),$$
gives the chemical potential or free energy per electron. This is 
the {\em entropic} part of the electrochemical potential (Fermi
energy) 
\begin{eqnarray}
\epsilon_{w_{i}} = \mu(n_i,T) - e \sum_{j=0}^{i-1}
(V_j + V_{w_{j}}) - {e V_{w_{i}}\over 2} .\label{elch}
\end{eqnarray}
According to (\ref{elch}), the Fermi energy, $\epsilon_{w_{
i}}$ (electrochemical potential), is the sum of the electrostatic
energy at the $i$th well, $-e\sum_{j=0}^{i-1} (V_{j}
+ V_{wj}) - e V_{wi}/2$, and the chemical potential, $\mu_i =
\mu(n_i,T)$.

After the change of variable in the integrals, the wave vectors
in (\ref{THM}) become:
\begin{eqnarray}
\hbar k_{i} = \sqrt{2 m_{w}^{*} \epsilon }\,,\nonumber \\
\hbar\alpha_{i} = \sqrt{2 m^{*} \left(e V_{b}
- {e V_{w_{i}}\over 2} -\epsilon\right)}\,, 
\nonumber\\
\hbar k_{i+1} = \sqrt{2 m_{w}^{*} \left(\epsilon + e V_{i} + e
{V_{w_{i}}+V_{w_{i+1}}\over 2}\right)}\,,\nonumber \\
\hbar\alpha_{i-1} = \sqrt{2 m^{*} \left(e V_{b} + {e V_{w_{i}}\over
2} + e V_{i-1} - \epsilon\right)}\,,\nonumber\\
\hbar\alpha_{i+1} = \sqrt{2 m^{*} \left(e V_{b}  - 
{e V_{w_{i}}\over 2} - e V_{i} - eV_{w_{i+1}} -
\epsilon\right)}\,,  \label{wavevec} 
\end{eqnarray}
where now $\epsilon=0$ at the bottom of the $i$th well.
This shows that the tunneling current density, $J_{i,i+1}$, in
(\ref{THM}) is a function of: the temperature, $\mu_i$
and $\mu_{i+1}$ (therefore of $n_i$ and $n_{i+1}$), the potential
drops $V_i$, $V_{i+1}$, $V_{w_{i}}$, and $V_{w_{i+1}}$:
\begin{eqnarray}
J_{i,i+1} =
\tilde{\Xi}(n_i,n_{i+1},V_i,V_{i+1},V_{w_{i}},V_{w_{i+1}}).
\label{tunnelfunction}
\end{eqnarray}
Similarly, we have
\begin{eqnarray}
J_{e,1} = \tilde{\Xi}_{e}(n_1,N_{D},V_{0},V_{w_{1}}),
\label{depJe}\\
J_{N,c} = \tilde{\Xi}_{c}(n_N,N_{D},V_{N},V_{w_{N}}).
\label{depJc}
\end{eqnarray}

\subsection{Balance and Poisson equations}
The 2D electron densities evolve according to the following rate
equations:
\begin{equation}
\frac{dn_{i}}{dt} = J_{i-1,i}-J_{i,i+1}
\hspace{2cm} i=1,\ldots,N.\label{rate}
\end{equation}
The voltage drops through the structure are calculated as
follows. The Poisson equation yields the potential drops in the
barriers, $V_{i}$, and the wells, $V_{wi}$ (see Fig.~\ref{fig1}):
\begin{eqnarray}
\varepsilon_{w}\,\frac{V_{w_{i}}}{w}
&=&\varepsilon\,\frac{V_{i-1}}{d}+\frac{e\,
( n_{i} - N_{D}^{w})}{2}\,,
\label{field.inside1}\\
\frac{V_{i}}{d}&=&\frac{V_{i-1}}{d}+\frac{e\, (n_{i}
-N_{D}^{w})}{\varepsilon}\,, \label{field.inside2}
\end{eqnarray}
where $\varepsilon_w$ and $\varepsilon$ are the GaAs and AlAs
static permittivities respectively,
$n_{i}$ is the 2D (areal) electron number density (to be
determined) which is singularly concentrated on a plane located at
the end of the $i$th well, and $N_{D}^{w}$ is the 2D intentional
doping at the wells.

\subsection{Boundary conditions}
The emitter and collector layers can be described by
the following equations:
\begin{eqnarray}
\frac{\varepsilon_{w}\Delta_{1}}{\delta_{1}} =
\frac{\varepsilon eV_{0}}{d}\,,\quad
\sigma_e = 2 \varepsilon\,\frac{V_{0}}{d}\simeq
eN(\epsilon_{F})\Delta_{1}\delta_{1}\,, \label{emitter}\\
\frac{\varepsilon_{w}\Delta_{2}}{e\delta_{2}}
=\frac{\varepsilon V_{N}}{d}-\frac{eN_{D}\delta_{2}}{2}
=  {\varepsilon_{w}\epsilon_{F}\over e\delta_{3}}\,,
\label{collector}\\
\sigma_c = 2 \varepsilon_w\,\frac{\epsilon_{F}}{e\delta_{3}}
= eN_{D}\left(\delta_{2}+{1\over 2}\delta_{3}\right)\, .
\label{charge.collector}
\end{eqnarray}
To write the emitter equations (\ref{emitter}), we assume that
there are no charges in the emitter barrier \cite{gol87}. Then 
the electric field  across $\delta_{1}$ (see Fig.~\ref{fig1}) is
equal to that in the emitter  barrier. Furthermore, the areal
charge density $\sigma_e$ required to create this electric field is
provided by the emitter. $N(\epsilon_{F}) = m_{w}^{*}\hbar^{-2} (3
N_D/\pi^4)^{{1\over  3}}$ is the density of states at the emitter
Fermi energy $\epsilon_{F}= \hbar^2 (3 \pi^2 N_D)^{{2\over
3}}/(2m_{w}^{*})$. The collector equations (\ref{collector}) and
(\ref{charge.collector}) ensure that the electrons tunneling
through the $N$th (last) barrier are captured by the collector.
They hold if the bias is large enough (see below). We assume that:
(i) the region of length $\delta_2$ in the collector is completely
depleted of electrons, (ii) there is local charge neutrality in the
region of length $\delta_3$ between the end of the depletion layer
$\delta_2$ and the collector, and (iii) the areal charge density
$\sigma_c$ required to create the local electric field is supplied
by the collector. Notice that $eN_{D}\left(\delta_{2}+{1\over
2}\delta_{3}\right)$ in (\ref{charge.collector}) is the positive
2D charge density depleted in the collector region. Equations
(\ref{collector}) and (\ref{charge.collector}) hold provided
$V_N\geq \varepsilon_w\epsilon_F d/(e\varepsilon\delta_3)$,
$\Delta_2\geq 0$,
$\delta_2\geq 0$ and $\delta_3\geq 0$. For smaller biases
resulting in $V_N<\varepsilon_w\epsilon_F
d/(e\varepsilon\delta_3)$, a boundary condition similar to
(\ref{emitter}) should be used instead of (\ref{collector}) and
(\ref{charge.collector}):
\begin{eqnarray}
\frac{\varepsilon_{w}\tilde{\Delta}_{2}}{\tilde{\delta}_{2}} =
\frac{\varepsilon eV_{N}}{d}\,,\quad \sigma_c = 2
\varepsilon\,\frac{V_{N}}{d}\simeq eN(\epsilon_{F})\tilde{
\Delta}_{2} \tilde{\delta}_{2}\,. \label{small} 
\end{eqnarray}
Notice that $\tilde{\Delta}_{2}$ and $\tilde{\delta}_{2}$ have
different meanings from $\Delta_{2}$ and $\delta_{2}$ in
(\ref{collector}). 

The condition of overall voltage bias closes the set of equations:
\begin{eqnarray}
V = \sum_{i=0}^{N}V_{i}+\sum_{i=1}^{N}V_{wi} +
\frac{\Delta_{1}+\Delta_{2}+\epsilon_{F}}{e} .\label{bias}
\end{eqnarray}
This condition holds only if $V_N \geq \varepsilon_w\epsilon_F
d/(e\varepsilon\delta_3)$; otherwise $(\Delta_2 + \epsilon_{F})$
should be replaced by
$\tilde{\Delta}_2$ in (\ref{bias}).

Notice that we can find $\delta_1$ and $\Delta_1$ as functions of
$V_0$ from (\ref{emitter}):
\begin{eqnarray}
\Delta_1 = 0 = V_0 ,\quad \delta_1\,\, \mbox{indetermined
or}\nonumber\\
\delta_1 = \sqrt{{2\varepsilon_{w}\over e^{2}N(\epsilon_{F})}}
= {\hbar \pi^{{2\over 3}}(2\varepsilon_{w})^{{1\over 2}}\over
e m_{w}^{*\, {1\over 2}} (3N_{D})^{{1\over 6}}}\,,\quad
\Delta_1 = {e\varepsilon V_{0}\delta_{1}\over
\varepsilon_{w} d}\,.\label{delta1}
\end{eqnarray}
Similarly we can find $\delta_3$ by solving (\ref{collector}) and
(\ref{charge.collector}) in terms of $V_N$ and $N_D$. From this
equation and (\ref{collector}), we can find $\delta_2$ and
$\Delta_2$ as functions of $V_N$:
\begin{eqnarray}
\delta_3 = {2\varepsilon \over eN_{D}d}\,\left[
\sqrt{V_{N}^{2}+{2\varepsilon_{w}\epsilon_{F}N_{D}d^{2}\over
\varepsilon^{2}}} - V_{N} \right]\,,\nonumber\\
\delta_2 = {2\varepsilon_{w}\epsilon_{F}\over e^{2}N_{D}
\delta_{3}}\,\left({eV_{N}\delta_{3}\varepsilon\over
\varepsilon_{w}\epsilon_{F}d} -
1\right)\,\theta\left({e\varepsilon V_{N}\delta_{3}\over
\varepsilon_{w}\epsilon_{F}d} - 1\right) ,\nonumber\\
\Delta_2 = {2\varepsilon\epsilon_{F}^{2}\over e^{2}N_{D}
\delta_{3}^{2}}\,\left({e\varepsilon V_{N}\delta_{3}\over
\varepsilon_{w}\epsilon_{F}d} - 1\right)\,
\theta\left({e\varepsilon V_{N}
\delta_{3}\over\varepsilon_{w}\epsilon_{F}d} -
1\right),\label{delta2}
\end{eqnarray}
where $\theta(x)$ is the Heaviside unit step function. The
boundary conditions (\ref{collector}) and (\ref{charge.collector})
do not hold if $e\varepsilon V_{N} \delta_3 <\varepsilon_{w}
\epsilon_F d$. This occurs if $V_N < (3\pi^2)^{{1\over 3}}
N_D^{{5\over 6}} \hbar d\sqrt{\varepsilon_{w}}/[2\varepsilon
\sqrt{2m_{w}^{*}}]$. In this case, we should impose the
alternative boundary conditions (\ref{small}). From these, we
obtain  
\begin{eqnarray} 
\tilde{\delta}_2 = \sqrt{{2\varepsilon_{w}\over
e^{2}N(\epsilon_{F})}} = {\hbar \pi^{{2\over
3}}(2\varepsilon_{w})^{{1\over 2}}\over e m_{w}^{*\, {1\over 2}}
(3N_{D})^{{1\over 6}}}\,,\nonumber\\ 
\tilde{\Delta}_2 = {e\varepsilon V_{N}\tilde{\delta}_{2}\over
\varepsilon_{w} d}\, \theta\left(1 - {e\varepsilon
V_{N}\delta_{3}\over\varepsilon_{w}\epsilon_{F}d}\right)\,. 
\label{solvedsmall}
\end{eqnarray}
The critical potential $V_N = \varepsilon_{w} \epsilon_{F}d/
(e\varepsilon\delta_{3})$ corresponds to $V_N = \varepsilon_{w}
\epsilon_{F}d/(\sqrt{3}\, e\varepsilon \tilde{\delta}_{2})$.
There is a small mismatch between (\ref{delta2}) and
(\ref{solvedsmall}) at this critical potential:
$e\varepsilon V_N/d = \varepsilon_{w}\epsilon_F/\delta_3$, but
$e\varepsilon V_N/d \neq \varepsilon_{w}
\epsilon_F/\tilde{\delta}_2$. This imperfection can be fixed by
using a more precise relation between the charge at the collector
$\sigma_c$, $\tilde{\Delta}_2$ and $\tilde{\delta}_2$, but we
choose not to delve more in these details. In all cases, we have
shown that the potential drops at the barriers separating the SL
from the contact regions uniquely determine the contact
electrostatics.

In Ref.\ \onlinecite{agu97} global charge conservation:
\begin{eqnarray}
\sigma_e +\sum_{i=1}^{N}(n_{i}-eN_{D}^{w}) = eN_{D}
(\delta_{2}+{1\over 2}\delta_{3}) \,,\label{charge.conserv}
\end{eqnarray}
was used instead of (\ref{charge.collector}) [which is a condition
similar to the one we impose at the emitter contact,
(\ref{emitter})]. Substitution of (\ref{charge.collector}) instead
of (\ref{charge.conserv}) modifies minimally the numerical results
reported in Refs.\ \onlinecite{agu97} and \onlinecite{san99}.

\subsection{Elimination of the potential drops at the wells}
The previous model has too many equations. We can eliminate
the potential drops at the wells from the system. For
(\ref{field.inside1}) and (\ref{field.inside2}) imply
\begin{eqnarray}
{\varepsilon_{w}V_{w_{i}}\over \varepsilon w} = {V_{i-1} +
V_{i}\over 2d}\,.\label{Vwell}
\end{eqnarray}
Then the bias condition (\ref{bias}) becomes
\begin{eqnarray}
V = \left(1+{\varepsilon w\over \varepsilon_{w}d}\right)\,
\sum_{i=0}^{N}V_{i} - {\varepsilon\, (V_{0}+V_{N})\, w\over
2\varepsilon_{w} d}\nonumber\\ 
+ \frac{\Delta_{1}+\Delta_{2}+\epsilon_{F}}{e}
\,,\label{bias2}
\end{eqnarray}
where $\Delta_1 = \Delta_1(V_0)$ and $\Delta_2 = \Delta_2(V_N)$.
Instead of the rate equations (\ref{rate}), we can derive a form of
Amp\`ere's law which explicitly contains the total current density
$J(t)$. We differentiate (\ref{field.inside2}) with respect to
time and eliminate $n_i$ by using (\ref{rate}). The result is
\begin{equation}
{\varepsilon\over e d}\frac{dV_{i}}{dt} + J_{i,i+1} = J(t),
\quad\quad\quad i=0,1,\ldots,N , \label{ampere}
\end{equation}
where $e J(t)$ is the sum of displacement and
tunneling currents. The time-dependent model consists of the
$2N+2$ equations (\ref{field.inside2}), (\ref{bias2}) and
(\ref{ampere}) [the currents are given by Eqs.~(\ref{THM}),
(\ref{n(e)}), (\ref{delta1}), (\ref{delta2}) and (\ref{Vwell})],
which contain the $2N+2$ unknowns $n_{i}$ ($j=1,\ldots,N$),
$V_j$ ($j=0,1,\ldots,N$), and $J$. Thus we have a system of
equations which, together with appropriate initial conditions,
determine completely and self-consistently our problem. For
convenience, let us list again the minimal set of equations we need
to solve in order to determine completely all the unknowns:
\begin{eqnarray}
{\varepsilon\over e d}\frac{dV_{i}}{dt} + J_{i,i+1} = J(t),
\quad\quad\quad i=0,1,\ldots,N ,\nonumber\\
\frac{V_{i}}{d} = \frac{V_{i-1}}{d}+\frac{e\, (n_{i}
- N_{D}^{w})}{\varepsilon}\,,\quad\quad i=1,\ldots,N ,
\label{mf2}\\
V = \left(1+{\varepsilon w\over \varepsilon_{w}d}\right)\,
\sum_{i=0}^{N}V_{i} - {(V_{0}+V_{N})\,\varepsilon w\over
2\varepsilon_{w}d}
\nonumber\\ + \frac{\Delta_{1}(V_{0})
+\Delta_{2}(V_{N})+\epsilon_{F}}{e}\,,\label{mf3}\\
J_{i,i+1} = \Xi(n_i,n_{i+1},V_{i-1},V_i,V_{i+1}),\label{mf4}\\
J_{e,1} = \Xi_e(n_{1},V_{0},V_1),\label{mf5}\\
J_{N,c} = \Xi_c(n_{N},V_{N-1},V_N). \label{mf6}
\end{eqnarray}

Notice that the three last equations are constitutive relations
obtained by substituting (\ref{Vwell}) in the functions $\tilde{\Xi}$,
$\tilde{\Xi}_e$ and $\tilde{\Xi}_c$ of (\ref{tunnelfunction}),
(\ref{depJe}) and (\ref{depJc}), respectively. The functions
$\Delta_1(V_0)$ and $\Delta_2(V_N)$ are given by (\ref{delta1}) and
(\ref{delta2}), respectively. Equations (\ref{ampere}) for $i=0,N$
may be considered the real boundary conditions for the barriers
separating the SL from the contacts. These boundary conditions are
the balance of current density including special tunneling
current constitutive relations $J_{e,1}$ and $J_{N,c}$. The latter
depend on the electron densities at the extreme wells of the SL
and the potential drops at the adjacent barriers.

\section{Derivation of the discrete drift-diffusion model}
\label{sec:3}
It is interesting to consider the relation (\ref{n(e)}) between the
chemical potential and the electron density
at a well for different temperature ranges:
\begin{eqnarray}
n_i(\mu_{i}) = \frac{m_{w}^{*} k_{B}T}{\pi^{2}\hbar^{2}}
\int A_{C1}(\epsilon) \ln
\left[1+e^{\frac{\mu_{i}-\epsilon}{k_{B}T}}\right]\,
d\epsilon . \nonumber
\end{eqnarray}
Assuming that $\mu_i\gg k_B T$, we may approximate this expression
by
\begin{eqnarray}
n_i(\mu_{i}) \approx \frac{m_{w}^{*}}{\pi^{2}\hbar^{2}}
\int_0^{\mu_{i}} A_{C1}(\epsilon)\, (\mu_{i}-\epsilon)\,
d\epsilon . \label{linear0}
\end{eqnarray}
Thus $n_i$ approaches a linear function of $\mu_i$ if $\mu_i\gg
k_B T$. For the SL used in the experiments we have been referring
to,  $\mu_i - \epsilon$ is typically about 20 meV or 232 K. Thus
``low temperature'' can be ``high enough temperature'' in
practice. Provided the Lorentzian $A_{C1}(\epsilon)$ is
sufficiently narrow, $A_{C1}(\epsilon)\sim \pi\,
\delta(\epsilon-\epsilon_{C1})$, so that
\begin{eqnarray}
\mu_{i}-\epsilon_{C1} \approx {\pi\hbar^{2}n_{i}\over
m_{w}^{*}}\quad
\mbox{if}\, (\mu_{i}-\epsilon_{C1})\gg k_{B} T,\,
\epsilon_{C1}\gg\gamma.
\label{T0}
\end{eqnarray}
Interestingly enough, a linear relation between $\mu_i$ and $n_i$
also holds at high temperatures. To derive it, notice that ln$(1+
e^x) \sim \ln 2 + x/2$ if $x\ll 1$ and use this relation in
(\ref{n(mu)}):
\begin{eqnarray}
n_i(\mu_{i}) \approx \frac{m_{w}^{*}}{2\pi^{2}\hbar^{2}}
\int_0^{\infty} A_{C1}(\epsilon)\, (2 k_B T\ln 2 + \mu_{i}
-\epsilon)\, d\epsilon. \label{linearinfty}
\end{eqnarray} 
If we now set $A_{C1}(\epsilon)\sim \pi\,
\delta(\epsilon-\epsilon_{C1})$, the result is 
$$\mu_{i}-\epsilon_{C1} \approx -2k_B T \ln2 + {2\pi\hbar^{2}
n_{i}\over m_{w}^{*}} 
$$ 
if $(\mu_{i}-\epsilon_{C1})\ll k_{B} T\ll
(\epsilon_{C2}-\epsilon_{C1})$, and $\epsilon_{C1}\gg\gamma$. The
additional condition (thermal energy small compared
to the difference between the energies of the two lowest
subbands) is needed to keep all electrons in the first subband.
For otherwise the second subband may be populated and Equation
(\ref{n(mu)}) should be transformed accordingly. Thus our ``high
temperature'' approximation can be satisfied in SL with large
enough energy differences $(\epsilon_{C2}-\epsilon_{C1})$.

A different approximation is obtained if we first impose that
$\epsilon_{C1}\gg\gamma$:
\begin{eqnarray}
n_i(\mu_{i})\approx \frac{m_{w}^{*} k_{B}T}{\pi\hbar^{2}}\, \ln
\left[1+e^{\frac{\mu_{i}-\epsilon_{C1}}{k_{B}T}}\right]\,.
\nonumber
\end{eqnarray}
This yields
\begin{eqnarray}
\mu_{i}\approx\epsilon_{C1}+ k_{B}T\, \ln
\left[e^{\frac{\pi\hbar^{2}n_{i}}{m_{w}^{*}k_{B}T}}-1\right]\,,
\nonumber
\end{eqnarray}
and therefore
\begin{eqnarray}
\mu_{i}-\epsilon_{C1} \approx {\pi\hbar^{2}n_{i}\over
m_{w}^{*}}\quad
\mbox{if}\quad \hbar^{2}n_{i}\gg m_{w}^{*} k_{B} T,\nonumber\\
\mu_{i}-\epsilon_{C1} \approx k_{B} T\,\ln{\pi\hbar^{2}n_{i}\over
m_{w}^{*}k_{B} T}\quad \mbox{if}\quad \hbar^{2}n_{i}\ll m_{w}^{*}
k_{B} T. \nonumber
\end{eqnarray}
At low temperatures, the chemical potential again depends linearly
on the electron density according to (\ref{T0}), whereas it has
ideal gas logarithmic dependence at high temperatures.

The same considerations used to obtain (\ref{linear0}) or
(\ref{linearinfty}) would indicate that the electron flux across
the $i$th barrier becomes
\begin{eqnarray}
J_{i,i+1}\approx {n_{i} v_{i}^{(f)} - n_{i+1} v_{i}^{(b)}
\over d+w}\,,\nonumber
\end{eqnarray}
either at low or high enough temperatures. Here $v_i^{(f)}$ and
$v_i^{(b)}$ are functions of $V_i$, $V_{i\pm 1}$. They have
dimensions of velocity and correspond to the forward and backward
tunneling currents which were invoked in the derivation of
phenomenological discrete drift models. When
$\epsilon_{w_{i}}= \epsilon_{w_{i+1}}$, or equivalently, $\mu_{i+1}
= \mu_i +eV_i + e(V_{w_{i}}+V_{w_{i+1}})/2$, $J_{i,i+1}=0$
according to (\ref{THM}). Equation (\ref{n(mu)}) implies that
$\mu_{i+1} = \mu_i$ if $n_{i+1} = n_i$, and therefore we conclude
that $v_{i}^{(f)} = v_{i}^{(b)}$ at zero potential drops $V_i +
(V_{w_{i}}+V_{w_{i+1}})/2 =0$. 
Notice that $\epsilon_{w_{i+1}} - \epsilon$ becomes $\mu_{i+1} - e
V_i - e V_{w_{i+1}}/2 - \epsilon'$ after changing variables in the
integral (\ref{THM}). Then $v_i^{(b)}$ is approximately zero unless
$0< \epsilon_{C1} + e V_{w_{i}}/2 < \mu_{i+1} - eV_i -
eV_{w_{i+1}}/2$. For voltages larger than those in the first
plateau of the current--voltage characteristic curve this
condition does not hold. In fact for these voltages, the level
$C1$ of well $i$ is at a higher or equal potential than the level
$C2$ of well $i+1$. Then $\epsilon_{C1}
\geq \mu_{i+1} - eV_i - e(V_{w_{i}}+V_{w_{i+1}})/2$.

The previous results yield DDD models with the potential drops at
the barriers and the total current density as unknowns, the same
as in Eqs.\ (\ref{ampere}) - (\ref{mf6}). The main difference with
previously used discrete drift models is that the velocity depends
on more than one potential drop. To obtain these simpler models,
we further assume that $\varepsilon V_i/\bar{\varepsilon} d$ and
$\varepsilon V_{i\pm 1}/\bar{\varepsilon}d$ are approximately
equal to an average field $F_i$ ($\bar{\varepsilon}$ is an average
permittivity to be chosen later). Then $V_{w_{i}} = w\bar{
\varepsilon} F_i/\varepsilon_w$ according to (\ref{Vwell}). This
assumption departs from previous approximations and yields a
new model. The point of contact with our previous results is that
$A_{C1}(\epsilon) A_{Cj}(\epsilon + eV_i+ e[V_{w_{i}} +
V_{w_{i+1}}]/2)$ is the controlling factor in the expressions for
$v^{(f)}$ and $v^{(b)}$ (the transmission coefficient contains an
exponential factor, $ e^{-2\alpha_{i}d}$, which is almost constant
at the energies contributing most to the integral). This
controlling factor is uniquely determined by the potential drop
$$V_i + {V_{w_{i}}+V_{w_{i+1}}\over 2} \approx
\left({d\over\varepsilon} + {w\over\varepsilon_{w}}\right)\,
\bar{\varepsilon}\, F_i =  (w+d)\, F_i ,$$
provided we define the average permittivity as
\begin{eqnarray} 
\bar{\varepsilon} = {d+w \over {d\over\varepsilon} +
{w\over\varepsilon_{w}}}\,.\label{bar}
\end{eqnarray} 
This expression corresponds to the equivalent capacitance of two
capacitors in series. Thus the behavior of forward and backward
drift velocities is most influenced by the potential drop $V_i+
(V_{w_{i}}+V_{w_{i+1}})/2 \approx F_i (d+w)$ and the new DDD model
(see below) should yield results similar to those of the
microscopic sequential tunneling model. We have 
\begin{eqnarray}  
J_{i,i+1}\approx {n_{i}
v^{(f)}(F_{i}) - n_{i+1} v^{(b)}(F_{i}) \over
d+w}\quad\quad\nonumber\\ = {n_{i} v(F_{i})\over d+w} -
{n_{i+1}-n_{i}\over (d+w)^{2}}\, D(F_{i}) ,\label{j1}\\ 
v(F) = v^{(f)}(F)- v^{(b)}(F),\,\, D(F) = (d+w)\, v^{(b)}(F). 
\label{j2}
\end{eqnarray}
To calculate $v^{(f)}(F)$ and $v^{(b)}(F)$ from $J_{i,i+1}$ in
(\ref{THM}), we replace $\epsilon_{w_{i}}$, $\epsilon_{C1}^i$, 
$\epsilon_{w_{i+1}}$ and $\epsilon_{Cj}^{i+1}$ by $\mu_i$,
$\epsilon_{C1}$, $\mu_{i+1} - e(d+w)F$ and $\epsilon_{Cj} -
e(d+w)F$, respectively. The wavevectors in the integrand should be 
\begin{eqnarray}
\hbar k_{i} = \sqrt{2 m_{w}^{*} \epsilon }\,,\nonumber \\
\hbar\alpha_{i} = \sqrt{2 m^{*} \left(e V_{b} -{ewF\over
2}- \epsilon\right)}\,, 
\nonumber\\
\hbar k_{i+1} = \sqrt{2 m_{w}^{*} [\epsilon + e (d+w)
F]}\,,\nonumber
\\
\hbar\alpha_{i-1} = \sqrt{2 m^{*} \left[e V_{b} + e
\left(d + {w\over 2}\right) F - \epsilon\right]}\,,\nonumber\\
\hbar\alpha_{i+1} = \sqrt{2 m^{*} \left[e V_{b} - e
\left(d+{3w\over   2} \right) F -
\epsilon\right]}\,,\label{wavevec1} 
\end{eqnarray}
and the integration variable $\epsilon$ ranges from 0 to $\infty$.
We substitute $\mu_i(n_i)$ according to (\ref{n(mu)}) in the
result. Then we obtain a function ${\cal J}(n_i,n_{i+1},F)$:
\begin{eqnarray}
{\cal J}(n_i,n_{i+1},F) =
\Xi\left(n_i,n_{i+1},{\bar{\varepsilon}Fd\over
\varepsilon}, {\bar{\varepsilon}Fd\over \varepsilon},
{\bar{\varepsilon}Fd\over\varepsilon}\right)\label{calJ}
\end{eqnarray}
[equivalent to setting $V_i =\bar{\varepsilon}F d/\varepsilon$,
or $V_i + (V_{w_{i}}+ V_{w_{i+1}})/2 = (d+w)\, F$ after
transforming this formula to the form (\ref{mf4})]. Notice that
(as said above)
$$v^{(f)}(0) = v^{(b)}(0) = {D(0)\over d+w}$$
for the tunneling current to vanish at zero field and equal
electron densities at adjacent wells. Furthermore, notice that
$D(F)$ vanishes if $\epsilon_{C1} - \mu_{i+1} \geq - e\, (2V_i +
V_{w_{i}}+V_{w_{i+1}})/2 \approx - e (d+w) F_i$. Thus according
to (\ref{T0}), $D(F)$ vanishes if $\hbar^2 n_{i+1} \leq
m_{w}^{*}e(d+w)F_i$, which is certainly satisfied for all average
fields larger than the first resonant field $(\epsilon_{C2}-
\epsilon_{C1})/[e(d+w)]$. In the low temperature limit (or in the
high temperature limit mentioned earlier in this Section,
provided it exists), we have 
\begin{eqnarray}
{\cal J}(n_i,n_{i+1},F) = {n_{i}\over d+w}\, v(F)
- {n_{i+1}-n_{i} \over (d+w)^{2}}\, D(F). \label{j3}
\end{eqnarray}
Then we may use
\begin{eqnarray}
v(F) &=&
{(d+w)\, {\cal J}(N_{D}^{w},N_{D}^{w},F)
\over  N_{D}^{w}}\,, \label{j4}\\ 
D(F) &=& - {(d+w)^{2}\, {\cal J}(0,N_{D}^{w},F)\over N_{D}^{w}}\,,
\label{j5}
\end{eqnarray}
to calculate the drift velocity and the diffusion coefficient from
the tunneling current. The integrals from (\ref{THM}) appearing in
these expressions may be approximated by means of the Laplace
method: we should just expand their controlling factor mentioned
before about its maximum value $\epsilon = \tilde{\epsilon}(F)$.
The resulting formulas are cumbersome and we choose not to write
them here. We show in the Appendix that $v^{(f)}(- F) = v^{(b)}(F)
\equiv D(F)/(d+w)$, and $v(-F) = - v(F)$. 

Equations (\ref{j3}) to (\ref{j5}) may be used in (\ref{ampere}) to
write the Amp\`ere law as
\begin{eqnarray}
{\bar{\varepsilon}\over e}\, {dF_{i}\over dt} + {n_{i}v(F_{i})\over
d+w} - D(F_i)\, {n_{i+1}-n_{i}\over (d+w)^{2}}= J(t)\,, \label{j6}
\end{eqnarray}
for $i=1,\ldots,N-1$. Poisson equation (\ref{mf2}) becomes
\begin{eqnarray}
F_{i}-F_{i-1} = {e\over\bar{\varepsilon}}\, (n_{i}-N_{D}^{w}) ,
\label{j7}
\end{eqnarray}
for $i=1,\ldots,N$. Equations (\ref{j6}) and (\ref{j7}) constitute
a DDD model which may be analyzed on its own together with
appropriate bias and boundary conditions. As bias condition we
adopt
\begin{eqnarray}
(d+w)\, \sum_{i=1}^{N} F_i = V\,.\label{j8}
\end{eqnarray}
Notice that potential drops at the contacts are represented only by 
the term $F_N\, (d+w)$. Equation (\ref{j8}) is obtained by
inserting $V_i+ (V_{w_{i}}+V_{w_{i+1}})/2 = (w+d) F_i$ into
(\ref{bias}), and omitting 
$$(d+w) F_{0} +\frac{\Delta_{1}+\Delta_{2}+2\epsilon_{F}}{2e}\,,
$$ 
for the sake of simplicity. For fields higher than the first
resonance, $D(F)\approx 0$, and (\ref{j6}) becomes
\begin{eqnarray}
{\bar{\varepsilon}\over e}\, {dF_{i}\over dt} + {n_{i}v(F_{i})\over
d+w} = J(t)\,, \label{drift}
\end{eqnarray}
which is the usual discrete drift model used in previous theoretical
studies \cite{bon94,bon95,kas97}.

\begin{figure}
\centerline{\epsfig{file=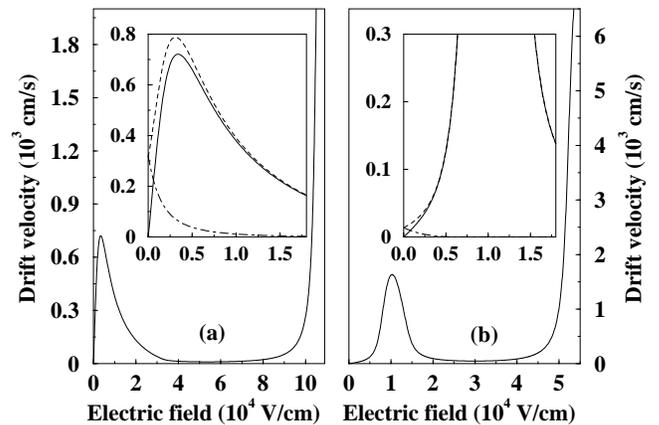,angle=270,width=0.45\textwidth}}
\vspace{0.5 cm}
\caption{(a) Electron drift velocity $v(F)$ for the 9/4 SL. Inset:
comparison of the drift velocity (continuous line) with the forward
(dashed line) and backward (dot-dashed line) velocities.
(b) The same for the
13.3/2.7 SL. Notice that the backward velocity or, equivalently
the diffusivity, decreases with electric field much more rapidly
for this SL.
}
\label{fig2}
\end{figure}

In Section 2.1 of Ref.\ \onlinecite{wac97a}, A.\ Wacker derived
a formula similar to (\ref{j1}) with $v^{(b)}=0$ and $v^{(f)}(F)\propto
\Gamma/[(eF(d+w)+\epsilon_{C1}-\epsilon_{Cj})^{2} + \Gamma^{2}]$,
for fields comparable to $(\epsilon_{Cj}-\epsilon_{C1})/[e(d+w)]$.
At low fields, the resonant tunneling current between levels $C1$ of
adjacent fields was found to be proportional to $W(F) = eF(d+w)/[(e^2
F^2 (d+w)^{2} + \Gamma_1^{2}]$ and independent of $n_i$. While the
first approximation of Wacker's (for fields close to higher resonances,
$C1\to Cj$, $j=2,3,\ldots$) is compatible with our result (\ref{j1}),
the second approximation is an artifact of the extra unnecessary assumption
$\epsilon_{w_{i}} = \epsilon_{w_{i+1}}$ \cite{wac97a}. We shall show
in Section \ref{sec:4} that our drift velocity (\ref{j4}) may have at
low fields the same shape as function $W(F)$ for certain SL; see
Fig.\ \ref{fig2}(a). Then the corresponding stationary current for a
space homogeneous field profile with $n_i = N_D^w$ (which implies equality
of chemical potentials at adjacent fields) will be proportional to $W(F)$
and our result will agree with Wacker's (for this special case). Fig.\
\ref{fig2}(b) shows that things may be different for a different SL
configuration.

The boundary conditions for $F_0$ and $F_N$ are also Amp\`ere's law
but now the tunneling currents (\ref{Je}) and (\ref{Jc}) (from the
emitter and to the collector, respectively) have to be used instead
of (\ref{THM}). The same approximations as before yield
\begin{eqnarray}
J_{e,1}&=& \Xi_{e}(n_1,\bar{\varepsilon}F_0 d/\varepsilon,
\bar{\varepsilon}F_0 d/\varepsilon)\nonumber\\ &\approx &
j_e^{(f)}(F_0) - {n_{1}\over d+w}\, w^{(b)}(F_{0})\,,
\label{j9}\\
J_{N,c}&=& \Xi_{c}(n_N,\bar{\varepsilon}F_N d/\varepsilon,
\bar{\varepsilon}F_N d/\varepsilon)\nonumber\\ 
&\approx & {n_{N}\over d+w}\, w^{(f)}(F_{N})\,.
\label{j10}
\end{eqnarray}
Notice that there is no backward tunneling from the collector
region to the SL because we are assuming that the potential drop
$V_N$ is larger than $\varepsilon_w\epsilon_F d/(e\varepsilon
\delta_3)$. Assuming now that (\ref{j9}) and (\ref{j10}) are
identities, we find
\begin{eqnarray}
j_e^{(f)}(F) &=&
\Xi_{e}\left(0,{\bar{\varepsilon}Fd\over\varepsilon},
{\bar{\varepsilon}Fd\over\varepsilon}\right),\label{j11}\\ 
w^{(b)}(F) &=&
{d+w\over N_{D}^{w}} \left[ j_e^{(f)}(F) -
\Xi_{e}\left(N_{D}^{w},{\bar{\varepsilon}Fd\over\varepsilon},
{\bar{\varepsilon}Fd\over\varepsilon}\right)\right],\label{j12}\\
 w^{(f)}(F) &=& {d+w\over N_{D}^{w}}\,
\Xi_{c}\left(N_{D}^{w},{\bar{\varepsilon}Fd\over\varepsilon},
{\bar{\varepsilon}Fd\over\varepsilon}\right)\,.\label{j14}
\end{eqnarray}
The tunneling current across a barrier is zero if the Fermi 
energies of the adjacent wells are equal. This occurs if the
electron density at the first well takes on an appropriate value
$n_{1}^{w}$ such that the corresponding Fermi energy equals that
of the emitter. Then
$$\Xi_{e}(n_{1}^{w},0,0) = 0, $$
and therefore
$$j_e^{(f)}(0) = {n_{1}^{w}\, w^{(b)}(0)\over d+w}\,. $$

\section{Numerical calculation of drift velocity and diffusion}
\label{sec:4}
In this Section, we shall calculate the functions $v(F)$, $D(F)$,
$j_e^{(f)}(F)$, $w^{(b)}(F)$ and $w^{(f)}(F)$ for different SL
used in experiments \cite{kas97}.
Fig.\ \ref{fig2}(a) depicts the electron drift velocity $v(F)$ 
for the 9nm/4nm GaAs/AlAs SL (9/4 SL) of Ref.\ \onlinecite{kas97}
calculated by means of (\ref{j4}) (at zero temperature; $m^{*} =
m_w^{*}$ for simplilcity). The inset compares $v(F)$ to the
backward and forward velocities given by
$v^{(b)}(F)=D(F)/(d+w)$ [$D(F)$ as in (\ref{j5})] and $v^{(f)}(F) =
v(F) + v^{(b)}(F)$. The rapidly decreasing diffusivity $D(F)$
determines the position and height of the first peak in $v(F)$.
Notice that $v(F)$ behaves as expected from general considerations:
it increases linearly for low electric fields, it reaches a maximum
and then decays before the influence of the second resonance is
felt. If $D(F)$ decays faster, a rather different $v(F)$ is found.
Fig.\ \ref{fig2}(b) shows $v(F)$ for the 13.3/2.7 SL: there is a
wide region before the first peak in which $v''(F)>0$.

Figures \ref{fig3} and \ref{fig4} show the boundary functions
$j_e^{(f)}(F)$, $w^{(b)}(F)$ and $w^{(f)}(F)$ for the 9/4 and
13.3/2.7 SL, respectively. Again they behave as expected: (i) the
emitter forward current peaks at the resonant values of the
electric field [different from those of $v^{(f)}(F)$], (ii) the
emitter backward tunnel velocity decreases rapidly with field, and
(iii) the collector forward velocity increases monotonically with
field given the large difference between the Fermi energies of the
last well and the collector.

The emitter forward current, $j_e^{(f)}(F)$, is compared in Figs.\
\ref{fig5} and \ref{fig6} to the drift current, $N_D^w v(F)/(d+w)$, for
different emitter doping values. Notice that the emitter current
is systematically higher than the drift current for large emitter
doping at positive electric fields. However, the total current
density should remain between the first maximum and the minimum of
the drift current. This means that the contact field $F_0$ should
be negative, so that $j_e^{(f)}(F_0) - n_{1}\, w^{(b)}(F_{0})/(d+w)
\approx J$, with $n_1>N_D^w$. Because of Poisson equation,
(\ref{mf2}), $F_1$ is larger than $F_0$ and, typically becomes
positive. The electric field in the SL increases with distance
from the emitter and a charge accumulation layer is formed (see
Figure 5 of Ref.\ \onlinecite{agu97} for a similar stationary field
profile corresponding to the full microscopic sequential tunneling
model). Self-consistent current oscillations in this situation will
be due to monopole recycling \cite{san99}. Notice that previous work
on discrete drift models assumed a fixed excess of electrons in the
first SL well as boundary condition \cite{bon95,kas97}. Again an
emitter accumulation layer appeared and monopole self-oscillation
resulted.

For smaller emitter doping, $j_e^{(f)}(F)$ intersects $N_D^w
v(F)/(d+w)$ on its second branch, and a charge depletion layer may
be formed in the SL. Then there may be self-oscillations due to
dipole recycling. These findings are fully consistent with the
numerical results reported in Ref.\ \onlinecite{san99} for the
13.3/2.7 SL. That paper reported coexistence and bistability of 
monopole and dipole self-oscillations for the first time.
Coexistence and bistability were found for an intermediate emitter
doping range (crossover range) {\em lower} than those used in
experiments \cite{san99}. A different way to obtain dipole
self-oscillations is to decrease the well width without changing
contact doping. In this way, we have numerically checked that
dipole self-oscillations are possible with emitter doping similar
to those used in current experimental setups
\cite{kas97}. 

For the usual drift-diffusion model of the Gunn effect
in bulk n-GaAs, the effect of boundary conditions on the
self-oscillations of the current has been well-studied
\cite{sha79,gom97}. In particular, asymptotic and numerical
calculations for realistic metal-semiconductor contacts were
performed some time ago \cite{gom97}. Despite the different
equations used in bulk semiconductors, these calculations provide
results consistent with our present findings in SL: a boundary
condition which yields accumulation (depletion) layer near
injecting contact may yield current self-oscillations due to
monopole (dipole) recycling \cite{gom97,sha79}. However these
similarities between discrete (SL) and continuous (bulk)
drift-diffusion models should not tempt us into reaching hasty
conclusions: discrete and continuous drift-diffusion models may
have rather different traveling wave solutions \cite{bon99}. In
fact, it has been shown that (depending on current and doping),
the DDD model may have monopole wave solutions which travel in the
same direction as the motion of electrons, in the opposite
direction, or remain stationary. In the continuum limit
(continuous drift-diffusion model), wavefronts travel always in
the same direction as the electrons \cite{bon99}. These features
of the DDD equations may have experimentally observable
consequences which will be explored elsewhere.

\section{Conclusions}
\label{sec:5}
Starting from a microscopic sequential tunneling model of transport
in weakly coupled SL, a DDD model is derived in the limits of low
or high temperature. Realistic transport coefficients and novel
contact current--field characteristic curves are calculated from
microscopic expressions, knowing the design parameters of the
superlattice. Boundary conditions select stable spatio-temporal
charge or field profiles in the SL. In prticular, they clarify when
possible self-sustained oscillations of the current are due to
monopole or dipole recycling. 

\acknowledgements
One of us (LLB) thanks Dr.\ Andreas Wacker for fruitful discussions
and collaboration on discrete drift-diffusion models. We thank
Dr.\ Ram\'on Aguado and Dr.\ Miguel Moscoso for fruitful
discussions. This work was supported by the Spanish DGES through
grants PB98-0142-C04-01 and PB96-0875, by the European Union TMR
contracts ERB FMBX-CT97-0157 and FMRX-CT98-0180 and by the
Community of Madrid, project 07N/0026/1998. 

\appendix
\section{Models for negative bias}
When a negative voltage is applied, we should make sure that our 
formulas transform appropriately. For negative bias, the
charge will be singularly concentrated on planes located at the 
beginning of the wells. Then we should write 
\begin{eqnarray}
\hbar \alpha_i = \sqrt{ 2 m^{*} \left[eV_{b} - e 
\sum_{j=0}^{i} (V_{j}+V_{w_{j}}) -\epsilon\right]},\nonumber
\end{eqnarray}
instead of (\ref{ai}) in the expressions (\ref{THM}). The change
of variable $\epsilon'=\epsilon + e W_{i+1}$ (i.e., $\epsilon'=0$ 
corresponds to zero energy at the bottom of well $i+1$) in the
integral (\ref{THM}), then changes the wavevectors to  
\begin{eqnarray} 
\hbar k_{i+1} = \sqrt{2 m_{w}^{*} \epsilon}\,, \nonumber \\ 
\hbar\alpha_{i} = \sqrt{2 m^{*} \left(e V_{b} +
{eV_{w_{i+1}}\over 2} -\epsilon\right)}\,,  \nonumber\\
\hbar k_{i} = \sqrt{2 m_{w}^{*} \left(\epsilon - e V_{i} - e
{V_{w_{i}}+V_{w_{i+1}}\over 2}\right)}\,,\nonumber \\
\hbar\alpha_{i-1} = \sqrt{2 m^{*} \left(e V_{b} + e V_{w_{i}} +
e V_{i} + {e V_{w_{i+1}}\over 2} - \epsilon\right)}\,,\nonumber\\
\hbar\alpha_{i+1} = \sqrt{2 m^{*} \left(e V_{b} - e V_{i+1} - {e
V_{w_{i+1}}\over 2} - \epsilon\right)}\,,\label{negwavevec} 
\end{eqnarray}
instead of (\ref{wavevec}). 

Given the new location of the singular charge planes (at the
beginning of wells), (\ref{field.inside1}) still holds, but
(\ref{field.inside2}) should be replaced by 
\begin{eqnarray}
\frac{V_{w_{i}}}{w}&=&\frac{V_{w_{i-1}}}{w}+\frac{e\, (
n_{i} - N_{D}^{w})}{\varepsilon_{w}}\,.
\label{neg1}
\end{eqnarray}
Then we find
\begin{eqnarray}
\frac{\varepsilon V_{i}}{\varepsilon_{w} d} =
\frac{V_{w_{i}}+V_{w_{i+1}}}{2w}\,, 
\label{neg2}
\end{eqnarray}
instead of (\ref{Vwell}). Inserting this equation in the functions
$\tilde{\Xi}$ (tunneling current under negative bias), we obtain
new functions
$\Xi^{*}(n_i,n_{i+1},V_{w_{i}},V_{w_{i+1}},V_{w_{i+2}})$, instead
of $\Xi(n_i,n_{i+1},V_{i-1},V_{i},V_{i+1})$ valid for positive
voltage. To obtain a reduced model, we now set
\begin{eqnarray}
\epsilon_{C1}^{i} = \epsilon_{C1}+e(d+w)F\,,\nonumber \\
\epsilon_{Cj}^{i+1} = \epsilon_{Cj}\,,\nonumber \\
\epsilon_{w_{i}} = \mu_{i}+e(d+w)F\,,\nonumber \\
\epsilon_{w_{i+1}} = \mu_{i+1}\,,\nonumber \\
\hbar k_{i+1} = \sqrt{2 m_{w}^{*} \epsilon }\,,\nonumber \\
\hbar\alpha_{i} = \sqrt{2 m^{*} \left(e V_{b} + {ewF\over
2} - \epsilon\right)}\,, 
\nonumber\\
\hbar k_{i} = \sqrt{2 m_{w}^{*} [\epsilon - e (d+w) F]}\,,\nonumber
\\
\hbar\alpha_{i-1} = \sqrt{2 m^{*} \left[e V_{b} + e
\left(d+{3w\over 2}\right) F - \epsilon\right]}\,,\nonumber\\
\hbar\alpha_{i+1} = \sqrt{2 m^{*} \left[ e V_{b} - e
\left(d+{w\over 2}\right) F - \epsilon\right]}\,,\label{wavevec2} 
\end{eqnarray}
in the integrals (\ref{THM}) and let the variable of integration
$\epsilon$ range from 0 to $\infty$. This is equivalent to setting 
$ V_{w_{i}}$, $V_{w_{i+1}}$ and $V_{w_{i+2}}$ equal to
$\bar{\varepsilon} w F/\varepsilon_w$ in
$\Xi^{*}(n_i,n_{i+1},V_{w_{i}},V_{w_{i+1}},V_{w_{i+2}})$.
Equations (\ref{THM}), (\ref{wavevec1}), (\ref{wavevec2}) and the
previous definitions in this Appendix imply 
\begin{eqnarray}
\Xi\left(n_i,n_{i+1},{\bar{\varepsilon}Fd\over \varepsilon},
{\bar{\varepsilon}Fd\over \varepsilon},{\bar{\varepsilon}Fd\over
\varepsilon}\right) =
\quad\quad\quad\quad \quad\quad
\quad\quad \nonumber\\ 
- \Xi^{*}\left(n_{i+1},n_i,-{\bar{\varepsilon}Fw\over
\varepsilon_{w}},-{\bar{\varepsilon}Fw\over
\varepsilon_{w}},-{\bar{\varepsilon}Fw\over
\varepsilon_{w}}\right). 
\label{neg3}
\end{eqnarray}
The Poisson equation (\ref{neg1}) still yields (\ref{j7}). Notice
that the symmetry (\ref{neg3}) implies 
\begin{eqnarray}
v^{(f)}(- F) = v^{(b)}(F) \equiv {D(F)\over d+w},
\quad v(-F) = - v(F).  
\label{neg4}
\end{eqnarray}

\begin{figure}
\centerline{\epsfig{file=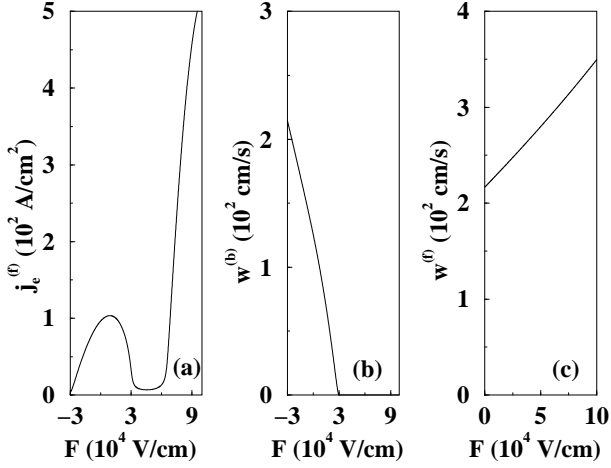,angle=270,width=0.45\textwidth}}		
\vspace{0.5 cm}
\caption{Functions of the electric field appearing in the boundary
conditions for the 9/4 SL with a contact doping $N_{D} = 2\times 10^{18}$
cm$^{-3}$. (a) $ej_e^{(f)}(F)$ and (b) $w^{(b)}(F)$ for the emitter and (c)
$w^{(f)}(F)$ for the collector. }
\label{fig3}
\end{figure}

\begin{figure}
\centerline{\epsfig{file=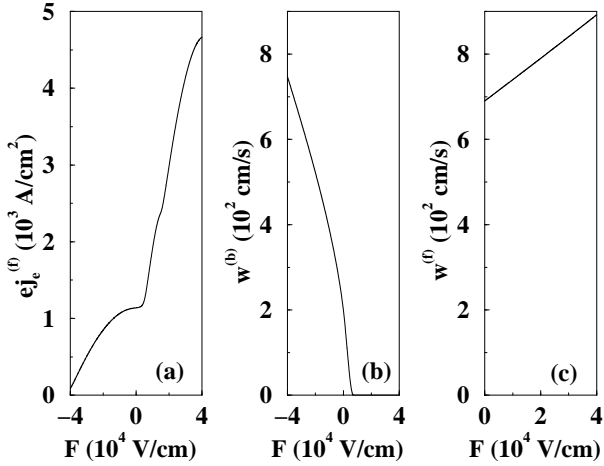,angle=270,width=0.45\textwidth}}
\vspace{0.5 cm}
\caption{Same functions as in Figure \ref{fig3} for the 13.3/2.7 SL
with a contact doping $N_{D} = 2\times 10^{18}$ cm$^{-3}$.
Notice that $ej_e^{(f)}(F)$ is an increasing function since
$\epsilon_{F}>(\epsilon_{C2}-\epsilon_{C1})$ in this SL. }
\label{fig4}
\end{figure}

\begin{figure}
\centerline{\epsfig{file=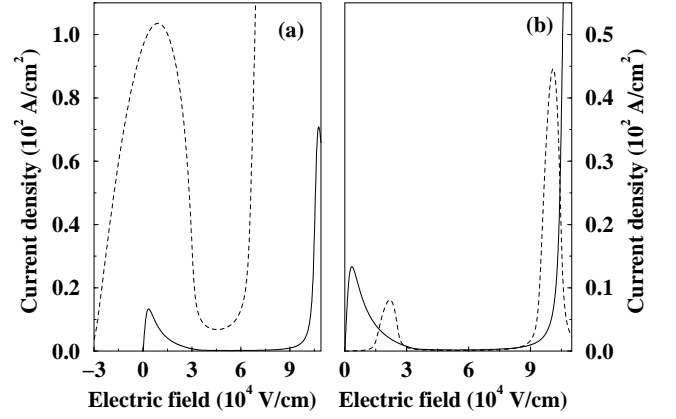,angle=270,width=0.45\textwidth}}
\vspace{0.5 cm}
\caption{Comparison of the drift tunneling current density, $eN_{D}^{w}\,
v(F)/(d+w)$ (continous lines) with the emitter current density $ej_e^{(f)}(F)$
(dashed lines) for the 9/4 SL
with two different emitter dopings:
(a) $N_{D} = 2\times 10^{18}$ cm$^{-3}$ corresponding to monopole recycling,
and (b) $N_{D} = 2\times 10^{17}$ cm$^{-3}$
corresponding to dipole recycling. }
\label{fig5}
\end{figure}

\begin{figure}
\centerline{\epsfig{file=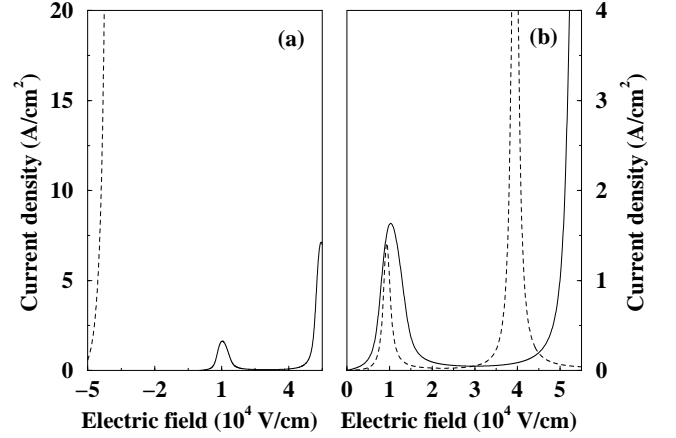,angle=270,width=0.45\textwidth}}
\vspace{0.5 cm}
\caption{Same functions as in Figure \ref{fig5} for the 13.3/2.7 SL
(a) $N_{D} = 2\times 10^{18}$ cm$^{-3}$ (monopole recycling),
and (b) $N_{D} = 10^{16}$ cm$^{-3}$ (dipole recycling). }
\label{fig6}
\end{figure}

Given the difference between the states at the contact regions and
the wells, the previous arguments cannot be used to extend the
contact current--field characteristic curves to negative fields.
Direct calculation of (\ref{j11}) - (\ref{j14}) by means of
(\ref{Je}) and (\ref{Jc}) yields the results depicted in Figures
\ref{fig3} and \ref{fig4}.


\end{multicols}

\end{document}